\documentclass[aps,superscriptaddress]{revtex4}

\usepackage[dvipdfm]{graphicx}
\usepackage{amsmath, amssymb}

\usepackage{color}

\usepackage{ulem}

\begin{document}
\title{Dispersal-induced destabilization of metapopulations and oscillatory Turing patterns in ecological networks} 
\author{Shigefumi Hata}
\affiliation{Department of Physical Chemistry, Fritz Haber Institute of the Max Planck Society, Faradayweg 4-6, 14195 Berlin, Germany.}
\author{Hiroya Nakao}
\affiliation{Department of Mechanical and Environmental Informatics, Tokyo Institute of Technology, Ookayama 2-12-1, 152-8552 Tokyo, Japan}
\author{Alexander S. Mikhailov}
\affiliation{Department of Physical Chemistry, Fritz Haber Institute of the Max Planck Society, Faradayweg 4-6, 14195 Berlin, Germany.}
\date{\today}
\maketitle


\section*{Abstract}
As proposed by Alan Turing in 1952 as a ubiquitous mechanism for nonequilibrium pattern formation,
diffusional effects may destabilize uniform distributions of reacting chemical species and lead to both spatially and temporally heterogeneous patterns.
While stationary Turing patterns are broadly known, the oscillatory instability,
leading to traveling waves in continuous media and also called the wave bifurcation, is rare for chemical systems.
Here, we extend the analysis by Turing to general networks and apply it to ecological metapopulations of biological species with dispersal connections between habitats.
Remarkably, the oscillatory Turing instability does not lead to wave patterns in networks, but to spontaneous development of heterogeneous oscillations and possible extinction of some species,
even though they are absent for isolated populations.
Furthermore, our theoretical analysis reveals that this instability is more common in ecological metapopulations than in chemical reactions.
Indeed, we find the instabilities for all possible food webs with three predator or prey species,
under various assumptions about the mobility of individual species and nonlinear interactions between them.
Therefore, we suggest that the oscillatory Turing instability is generic and must play a fundamental role in metapopulation dynamics,
providing a common mechanism for dispersal-induced destabilization of ecosystems.


\section{Introduction}

A rich variety of nonequilibrium pattern formation is supported by reaction-diffusion processes.
One of the universal mechanisms of such pattern formation is provided by the Turing instability~\cite{Turing1952};
a diffusion-induced instability of the homogenous state which leads to spontaneous development of self-organized patterns.
The Turing instability can play an important role in biological morphogenesis and it has been extensively studied in various applications, including biological~\cite{Meinhardt2000, Murray2003,Maini2006,Sick2006,Kondo1995}, chemical~\cite{Castets1990,Ouyang1991} and physical systems~\cite{Nakao2010}.
The classical Turing instability leads to the establishment of stationary spatial patterns.
However, the oscillatory analogue of this instability is possible and it has also been discovered by Alan Turing~\cite{Turing1952}.
This oscillatory intability produces traveling or standing waves and therefore it is often called ``the wave instability'' (see~\cite{Walgraef1997}).
He has shown that at least three species are needed for the oscillatory instability,
while the stationary instability is possible already with two species.
The stationary Turing instability has been extensively studied both theoretically~\cite{Meinhardt2000,Murray2003,Maini2006,Nakao2010}and experimentally~\cite{Sick2006,Kondo1995,Castets1990,Ouyang1991},
whereas the oscillatory instability is more rare and it was found only for special chemical systems~\cite{Zhabotinsky1995, Vanag2001, Yang2002}.
Note that complex spatio-temporal patterns can also emerge as a result of interactions between the stationary Turing instability and other bifurcations.
For example, near the Turing-Hopf bifurcation point, complex mixed modes leading to standing waves and spatio-temporal chaos can exist~\cite{DeWit1996}.
However, their mechanism is different from that of the oscillatory Turing instability (or the wave bifurcation).

Reaction-diffusion processes are also characteristic to ecological systems.
The reactions correspond in this case to the predator-prey and other interactions between the species.
Both passive diffusion and active random migration are possible in ecological populations.
Moreover, there are situations when a population occupies a large habitat and therefore can be considered as an extended spatial system.
The classical Turing instability is possible in ecosystems.
It has been shown that such instability should be generic for the two-species predator-prey models~\cite{Baurmann2007} (see also~\cite{Yi2009}).
Complex spatio-temporal ecological dynamics related to the Turing-Hopf~\cite{Baurmann2007, Melese2011}
and Turing-Takens-Bogdanov~\cite{Baurmann2007} bifurcations has been discussed.
Stationary Turing patterns have been found in realistic models describing plant-parasite~\cite{White1998}, plankton-fish~\cite{Medvinsky2002} and plant-insect~\cite{DellaRossa2012} interactions.
The oscillatory Turing instability is also possible in ecology.
In a study of a three-species plant-parasite-hyperparasite system,
such instability leading to standing waves, has been previously considered~\cite{White1998}.

While some ecological systems can be described by reaction-diffusion equations for continuous media,
there are also many ecosystems that are spatially fragmented and represent networks~\cite{Urban2001, Fortuna2006, Minor2008}.
Such networks are formed by individual habitats which are linked by dispersal connections.
Ecological species populate the habitats and diffusively migrate over a network.
Such network-organized ecosystems are known as metapopulations~\cite{May1974, Hanski1991, Hanski1998}.
The metapopulation concept has been applied to describe and investigate real ecological systems (see, e.g.~\cite{Gonzalez1998, Fahrig2003, Keymer2006}).
It has also been used in the context of the epidemic research~\cite{Satorras2001, Hufnagel2004, Colizza2006, Colizza2008}.
In the framework of the metapopulation concept, the role of dispersal connections in the stability enhancement of an ecosystem ({\it the rescue effect}) has been discussed~\cite{Hanski1991} (see also~\cite{Fahrig2003}).
The theoretical results have been tested in the experiments with specially prepared metapopulations~\cite{Gonzalez1998, Keymer2006}.

Ecological metapopulations provide examples of reaction-diffusion systems with a network structure.
Such systems can however be also found in other research fields.
For instance, a biological embryo can be viewed as a network of cells with the chemicals diffusing over the pattern of intercellular connections~\cite{Othmer1971, Othmer1974, Bignone2001}.
Networks formed by coupled reactors can also be considered~\cite{Horsthamke2004, Moore2005}.
Theoretical studies of reaction-diffusion processes on networks have already attracted much attention~\cite{Barrat2008}.
Effects of infection spreading over the networks have been discussed in detail~\cite{Satorras2001, Hufnagel2004, Colizza2006, Colizza2008}.
The role of network topology on the phase diagrams of nonequilibrium phase transitions on networks has been considered~\cite{Colizza2007}.
Traveling and pinned fronts in networks of diffusively coupled bistable elements were analyzed~\cite{Kouvaris2012} and control of front propagation by global feedback has been considered~\cite{Kouvaris2013}.
There is large literature on the networks formed by diffusively coupled oscillators~\cite{Boccaletti2006, Arenas2008} (see also~\cite{Nakao2009}).
The role of dispersal connections in the synchronization effects in ecological networks has also been discussed~\cite{Holland2008}.

The stationary Turing instability for networks has been first analyzed in 1971 by Othmer and Scriven~\cite{Othmer1971}.
The authors have introduced a general mathematical description
of the classical Turing instability in two-component reaction-diffusion networks
and have applied their theory for regular lattices (see also~\cite{Othmer1974}).
Stationary Turing patterns in small networks of coupled chemical reactors
have subsequently been discussed~\cite{Horsthamke2004}.
The properties of such instability and of the final established stationary patterns in large random networks of diffusively coupled activator-inhibitor elements have been investigated
and the mean-field theory of Turing patterns in such network systems has been constructed~\cite{Nakao2010}.
The global feedback control of the stationary Turing patterns in networks has been studied~\cite{Hata2012}.
A detailed mathematical analysis of the hysteresis phenomena related to the network Turing bifurcation has recently been performed~\cite{Wolfrum2012}.

In the present article, we provide, for the first time, a general mathematical theory of the oscillatory Turing instability (the analogue of the wave bifurcation) in network-organized reaction-diffusion systems
and apply the theory to large ecological networks.
Our numerical simulations are performed for the metapopulations with various three-species food webs
under different assumptions about the nonlinearities of the population dynamics.
The instability could be found for all such systems and, as we therefore believe,
it should be common in ecology.
In contrast to the wave instability in continuous media, traveling or standing waves do not develop
and oscillations, localized on a subset of network nodes, are instead observed.
The bifurcation is always supercritical and therefore the final pattern is usually well described by the first critical mode.
This diffusion-induced instability leads to destabilization of metapopulations and the extinction of some species may be its result.

This article is organized as follows:
In section~\ref{sec_model}, the addressed problem is mathematically formulated.
General descriptions for ecological networks are introduced and specific mathematical models which we employ are explained.
In section~\ref{sec_analytic}, analytical investigations, including
the linear stability analysis for general reaction-diffusion networks with three reacting species, is undertaken.
Sufficient conditions for the oscillatory Turing instability in ecological systems are constructed.
Numerical investigations for various ecological models are performed in section~\ref{sec_num}.
First, examples of oscillatory and stationary patterns in networks are demonstrated.
After that, a detailed analysis of the simulation results is performed.
As we find, the oscillatory bifurcation is always supercritical and the small-amplitude patterns are well described by the critical modes which correspond to certain Laplacian eigenvectors.
The localization of the Laplacian eigenvectors for networks explains the observed localization of the oscillatory Turing patterns on a subset of network nodes.
An example of a secondary instability of the oscillatory patterns, leading to the extinction of one of the species, is given.
Finally, we summarizes the results and discuss them in Section~\ref{sec_sum}.
In Appendices, numerical simulations for additional ecological models and for two chemical network models are reported.

\newpage
\section{Formulation of the problem}
\label{sec_model}
\begin{figure}[tb]
\includegraphics[width=84mm]{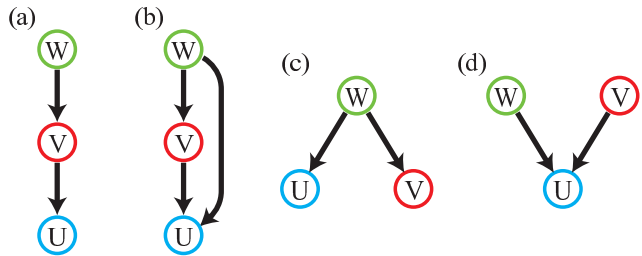}
\caption{
Food web diagrams.
Each arrow goes from the consuming to the consumed species.
In the last food web (d), no persistence with fixed values of $(\bar u, \bar v, \bar w)$ can be achieved;
therefore this web is excluded from our analysis.
}
\label{fig01}
\end{figure}

We consider ecological networks formed by individual populations which occupy separate habitats and are coupled by dispersal connections.
Our attention is focused on the populations which consist of three interacting species.
All possible food webs with three different species are displayed in Fig.~\ref{fig01}.
Generally, such ecological networks are described by equations
\begin{align}
\begin{cases}
\displaystyle \frac{du_{i}}{dt} =  F\left ( u_{i}, v_{i}, w_{i} \right ) u_i+ \epsilon \sigma_{u} \displaystyle \sum_{j=1}^N L_{ij} u_j,\\
\displaystyle \frac{dv_{i}}{dt} =  G\left ( u_{i}, v_{i}, w_{i} \right ) v_i+ \epsilon \sigma_{v} \displaystyle \sum_{j=1}^N L_{ij} v_j,\\
\displaystyle \frac{dw_{i}}{dt} =  H\left ( u_{i}, v_{i}, w_{i} \right ) w_i+ \epsilon \sigma_{w} \displaystyle \sum_{j=1}^N L_{ij} w_j,
\label{eq01}
\end{cases}
\end{align}
for $i,j=1,\cdots,N$, 
where population densities of species on node $i$ are denoted as $u_{i} = [U]_{i}, v_{i} = [V]_{i}$, and $w_{i} = [W]_{i}$,
functions $F=Q^{u}-R^{u}, G=Q^{v}-R^{v}, H=Q^{w}-R^{w}$ are the differences of reproduction ($Q$) and death ($R$) rates for each species,
and $\sigma_{u,v,w}$ are the mobilities of the three species;
the common parameter $\epsilon$ is introduced for convenience, so that the mobility of all species can be varied without changing relative mobilities.
The Laplacian matrix $\mathbf{L}$ has elements $L_{ij} = A_{ij}-\sum_{j} A_{ij}\delta_{ij}$ where $A_{ij}$ is the matrix of connections between the habitats.
For simplicity, we assume that all connections, if present, have the same strength.
Therefore, we have $A_{ij}=1$, if there is a connection between habitats $i$ and $j$, and $A_{ij}=0$ otherwise.
We assume that, in absence of diffusive coupling, a stable stationary state $(u_{0},v_{0},w_{0})$ exists
which is determined by $F(u_{0},v_{0},w_{0})=G(u_{0},v_{0},w_{0})=H(u_{0},v_{0},w_{0})=0$
where $u_{0}>0, v_{0}>0$ and $w_{0}>0$.

We examine three different food webs shown in Fig.~\ref{fig01}(a)-(c).
Below, we provide explicit expressions for the reproduction rates and death rates in each model.
While different nonlinear dependence can be considered for predator-prey interactions,
we will employ the Holling type II functions~\cite{Holling1959,Murray2003}
(See Supporting Information for other dependences).
Note that the food web shown in Fig.~\ref{fig01}(d) is excluded from our analysis.
In such system, two competing species $V$ and $W$ cannot coexist in a steady uniform state
and Turing instabilities with three reacting species are impossible.\\
{\bf Model A.}
In the food chain shown in Fig.~\ref{fig01}(a), top predator $W$ is feeding on intermediate species $V$ which is in turn a predator for prey $U$.
The collective dynamics of such metapopulation is described by Eqs.~(\ref{eq01}) with
\begin{align}
& Q^{u}= a_{u},&\quad &R^{u}(u,v)=b_{u}u + c_{u}\frac{v}{u+\mu},\cr
& Q^{v}(u)=c_{v}\frac{u}{u+\mu},&\quad &R^{v}(v,w)= a_{v} + d_{v}\frac{w}{v+\nu},\label{eq02}\\
& Q^{w}(v)=d_{w}\frac{v}{v+\nu},& \quad &R^{w}= a_{w},\nonumber
\end{align}  
{\bf Model B.}
In the food web shown in Fig.~\ref{fig01}(b), both species $V$ and $W$ play a role of the predators for prey $U$ while $V$ is also a prey for $W$.
Such system is modelled as follows:
\begin{align}
&Q^{u}(u) = a_{u} - b_{u}u,& &R^{u}(u,v,w) = c_{u}\frac{v}{u+\mu}+e_{u}\frac{w}{u+\rho},\cr
&Q^{v}(u) = c_{v}\frac{u}{u+\mu},& &R^{v}(v,w) = a_{v} + d_{v}\frac{w}{v+\nu},\label{eqS06}\\
&Q^{w}(u,v) = d_{w}\frac{v}{v+\nu} + e_{w}\frac{u}{u+\rho},& &R^{w} = a_{w}.\nonumber
\end{align}
{\bf Model C.}
In the food web shown in Fig.~\ref{fig01}(c), species $U$ and $V$ are the prey for predator $W$.
Thus we have
\begin{align}
&Q^{u}(u) = a_{u} -b_{u}u,& &R^{u}(u,w) = c_{u}\frac{w}{u+\mu},\cr
&Q^{v}(v) = a_{v} - b_{v}v,& &R^{v}(v,w) = d_{v}\frac{w}{v+\nu},\label{eqS07}\\
&Q^{w}(u,v) =  c_{w}\frac{u}{u+\mu} + d_{w}\frac{v}{v+\nu},& &R^{w} = a_{w}.\nonumber
\end{align}

For comparison, the continuous analog of the model (Eqs.~(\ref{eq01})) will be also considered.
Then, node variables $u_{i},v_{i}$ and $w_{i}$ are replaced by space-dependent densities $u,v$ and $w$
and differential diffusion terms like $\epsilon' \sigma_{u} \nabla^{2} u$ are introduced instead of the last terms in Eqs.~(\ref{eq01}).

\section{Analytical investigations}
\label{sec_analytic}
\subsection{Linear stability analysis}
\label{sec_linear_stab}
Below we give the precise definition of the oscillatory Turing instability in ecological networks.
The stability of the uniform state of the model can be analyzed by the linear stability analysis.
Small perturbations $(\delta u_{i},\delta v_{i},\delta w_{i})$ are introduced to the steady state as
$(u_{i}, v_{i}, w_{i}) = (u_0, v_0, w_0) + (\delta u_{i},\delta v_{i},\delta w_{i})$.
Substituting this into Eqs.~(\ref{eq01}),
the following linearized differential equations are obtained:
\begin{align}
\begin{cases}
\displaystyle \frac{d}{dt}\delta u_i = F_{u}u_0\delta u_{i} + F_{v}u_0\delta v_{i} + F_{w}u_0\delta w_{i} + \epsilon \sigma_{u} \displaystyle \sum_{j=1}^N L_{ij} \delta u_j,\\
\displaystyle \frac{d}{dt}\delta v_i = G_{u}v_0\delta u_{i} + G_{v}v_0\delta v_{i} + G_{w}v_0\delta w_{i} + \epsilon \sigma_{v} \displaystyle \sum_{j=1}^N L_{ij} \delta v_j,\\
\displaystyle \frac{d}{dt}\delta w_i = H_{u}w_0\delta u_{i} + H_{v}w_0\delta v_{i} + H_{w}w_0\delta w_{i} + \epsilon \sigma_{w} \displaystyle \sum_{j=1}^N L_{ij} \delta w_j.
\label{eq06}
\end{cases}
\end{align}
where $F_u = \partial F / \partial u |_{(u_0, v_0, w_0)}$,
$F_v = \partial F / \partial v |_{(u_0, v_0, w_0)}$,
$F_w = \partial F / \partial w |_{(u_0, v_0, w_0)}$,
... are partial derivatives at the uniform steady state $(u_0, v_0, w_0)$.
The Laplacian eigenvectors $\{{\vec \phi}^{(\alpha)} \}$ are introduced to decompose the perturbations.
They are defined as
\begin{align}
\sum_{j=1}^{N} L_{ij} \phi_{j}^{(\alpha)} = \Lambda^{(\alpha)}\phi^{(\alpha)}_{i}
\label{eq07}
\end{align}
where $\Lambda^{(\alpha)}$ is the Laplacian eigenvalue of the $\alpha$ th mode ($\alpha=1,\cdots,N$).
The mode indices $\{\alpha\}$ are sorted in the increasing order of the Laplacian eigenvalues $\{ \Lambda^{(\alpha)} \}$ so that
$\Lambda^{(1)}\leq \Lambda^{(2)}\leq \cdots \leq \Lambda^{(N)} = 0$ holds.
The perturbations $(\delta u_{i}, \delta v_{i}, \delta w_{i})$ are expanded over the set of the Laplacian eigenvectors as
\begin{align}
\delta u_{i}(t) &= \sum_{\alpha=1}^{N} U^{(\alpha)}\exp [\gamma^{(\alpha)} t] \phi^{(\alpha)}_{i}\cr
\delta v_{i}(t) &= \sum_{\alpha=1}^{N} V^{(\alpha)}\exp [\gamma^{(\alpha)} t] \phi^{(\alpha)}_{i}\label{eq08}\\
\delta w_{i}(t) &= \sum_{\alpha=1}^{N} W^{(\alpha)}\exp [\gamma^{(\alpha)} t] \phi^{(\alpha)}_{i},\nonumber
\end{align}
where $\gamma^{(\alpha)}=\lambda^{(\alpha)}+i\omega^{(\alpha)}$ is a complex growth (or decay) rate of the $\alpha$ th eigenmode.
Substituting these expressions into Eqs.~(\ref{eq06}), the following equations are obtained for each eigenmode:
\begin{equation}
\gamma^{(\alpha)}
\begin{pmatrix}
U^{(\alpha)}\\
V^{(\alpha)}\\
W^{(\alpha)}
\end{pmatrix}
=
\begin{pmatrix}
F_u u_0+\epsilon \sigma_{u} \Lambda^{(\alpha)} & F_v u_0 & F_{w} u_0\\
G_u v_0& G_v v_0+ \epsilon \sigma_{v} \Lambda^{(\alpha)} & G_{w}v_0\\
H_{u} w_0& H_{v}w_0 & H_{w}w_0 + \epsilon \sigma_{w} \Lambda^{(\alpha)}
\end{pmatrix}
\begin{pmatrix}
U^{(\alpha)}\\
V^{(\alpha)}\\
W^{(\alpha)}
\end{pmatrix}.
\label{eq09}
\end{equation}
Thus, the decay rate of each eigenmode is determined by the characteristic equation
\begin{equation}
\textrm{Det}
\begin{pmatrix}
F_u u_0+\epsilon \sigma_{u} \Lambda^{(\alpha)}-\gamma^{(\alpha)} & F_v u_0 & F_{w} u_0\\
G_u v_0& G_v v_0+ \epsilon \sigma_{v} \Lambda^{(\alpha)}-\gamma^{(\alpha)} & G_{w}v_0\\
H_{u} w_0& H_{v}w_0 & H_{w}w_0 + \epsilon \sigma_{w} \Lambda^{(\alpha)}-\gamma^{(\alpha)}
\end{pmatrix}
=0.
\label{eq10}
\end{equation}
The uniform steady state is stable if $\lambda^{(\alpha)}<0$ for all $\alpha$.
The Turing instability takes place if $\lambda^{(\alpha)}$ becomes positive at some $\alpha = \alpha_{c}$ which represents the critical mode for the instability.
The critical modes are stationary, $\phi_{i}^{(\alpha_{\textrm c})}$, if $\omega^{(\alpha_{\textrm c})} = 0$.
On the other hand, the critical modes can also be oscillatory, $\phi_{i}^{(\alpha_{\textrm c})}e^{i\omega^{(\alpha_{\textrm c})} t}$, if $\omega^{(\alpha_{\textrm c})} \neq 0$.
As noticed already by Turing~\cite{Turing1952}, oscillatory instabilities are possible only if the number of species is at least three.

In the continuous case, the diffusion operator is $\nabla^{2}$ and its eigenvectors are plane waves,
since $\nabla^{2}e^{ikx}=-k^{2}e^{ikx}$.
Thus, small perturbations are decomposed over plane waves as
\begin{align}
\delta u(x,t) &= \int dk U^{(k)}\exp [\gamma^{(k)} t] \exp[ikx]\cr
\delta v(x,t) &= \int dk V^{(k)}\exp [\gamma^{(k)} t] \exp[ikx]\label{eq11}\\
\delta w(x,t) &= \int dk W^{(k)}\exp [\gamma^{(k)} t] \exp[ikx]\nonumber
\end{align}
to obtain the characteristic equation
\begin{equation}
\textrm{Det}
\begin{pmatrix}
F_u u_0-\epsilon ' \sigma_{u} k^2-\gamma^{(k)} & F_v u_0 & F_{w} u_0\\
G_u v_0& G_v v_0- \epsilon ' \sigma_{v} k^2-\gamma^{(k)} & G_{w}v_0\\
H_{u} w_0& H_{v}w_0 & H_{w}w_0 - \epsilon ' \sigma_{w} k^2-\gamma^{(k)}
\end{pmatrix}
=0.
\label{eq12}
\end{equation}
The instability occurs if $\lambda^{(k)}$ becomes positive at some critical wave number $k_{\textrm c}$.
The critical mode corresponds to a traveling wave $\phi_{i}^{(k_{\textrm c})}e^{i\omega^{(k_{\textrm c})} t}$ if $\omega^{(k_{\textrm c})} \neq 0$ or to a stationary wave $e^{ik_{\textrm c}x}$ if $\omega^{(k_{\textrm c})} = 0$.

For continuous media, the oscillatory Turing instability is usually called the wave instability,
since the first unstable modes are the traveling waves.
In networks, traveling waves do not appear
and therefore it is not appropriate to talk about a wave instability in this case. 

\subsection{Sufficient conditions for the oscillatory Turing instability in ecological networks}
\label{sec_sufficient}
In our recent work~\cite{Hata_arXiv}, we considered general continuous reaction-diffusion systems with three species
and derived sufficient conditions for the oscillatory Turing instability in continuous media.
The derivation was based on the linear stability analysis.
This analysis can be straightforwardly extended to the case of discrete networks.
Below, we only reformulate the previously derived sufficient conditions for the networks
and apply them to the considered ecological models.

Suppose that the mobilities $\sigma_{u}$ and $\sigma_{v}$ of the species $U$ and $V$ are fixed
(and at least one of them is non-vanishing)
and we gradually increase the mobility $\sigma_{w}$ of species $W$.
It follows from our previous analysis~\cite{Hata_arXiv} that the oscillatory Turing instability will always be found at sufficiently high mobility $\sigma_{w}$ if the following three sufficient conditions are satisfied:
\begin{subequations}
\begin{align}
& F_{u}u_0+G_{v}v_0>0,
\label{eqA16a}\\
& F_{w}H_{u}u_{0}+G_{w}H_{v}v_{0}<0,
\label{eqA16b}\\
& \textrm{Det}
\begin{pmatrix}
F_u u_0+\sigma_{u} \Lambda & F_v u_0\\
G_u v_0& G_v v_0+ \sigma_{v} \Lambda
\end{pmatrix}
\neq 0 \ \textrm{for any} \ \Lambda<0.
\label{eqA16c}
\end{align}
\end{subequations}
Note that one of the species $U$ or $V$ may be immobile, i.e. either $\sigma_{u}=0$ or $\sigma_{v}=0$.

If the mobilities $\sigma_{v}$ and $\sigma_{w}$ of the species $V$ and $W$ are fixed
(and at least one of them is non-vanishing)
and we gradually increase the mobility $\sigma_{u}$ of species $U$,
the oscillatory Turing instability will always be found at sufficiently high mobility $\sigma_{u}$ if the following three sufficient conditions are satisfied:
\begin{subequations}
\begin{align}
& G_{v}v_0+H_{w}w_0>0,
\label{eqA17a}\\
& G_{u}F_{v}v_{0}+H_{u}F_{w}w_{0}<0,
\label{eqA17b}\\
& \textrm{Det}
\begin{pmatrix}
G_v v_0+\sigma_{v} \Lambda & G_w v_0\\
H_v w_0& H_w w_0+ \sigma_{w} \Lambda
\end{pmatrix}
\neq 0 \ \textrm{for any} \ \Lambda<0.
\label{eqA17c}
\end{align}
\end{subequations}
Here, one of the species $V$ or $W$ may be immobile so that either $\sigma_{v}=0$ or $\sigma_{w}=0$.

When the mobilities $\sigma_{w}$ and $\sigma_{u}$ of the species $W$ and $U$ are instead fixed
(and at least one of them is non-vanishing)
and we gradually increase the mobility $\sigma_{v}$ of species $V$,
the oscillatory Turing instability will always be found at sufficiently high mobility $\sigma_{v}$ if the following three sufficient conditions are satisfied:
\begin{subequations}
\begin{align}
& H_{w}w_0+F_{u}u_0>0,
\label{eqA18a}\\
& H_{v}G_{w}w_{0}+F_{v}G_{u}u_{0}<0,
\label{eqA18b}\\
& \textrm{Det}
\begin{pmatrix}
H_w w_0+\sigma_{w} \Lambda & H_u w_0\\
F_w u_0& F_u u_0+ \sigma_{u} \Lambda
\end{pmatrix}
\neq 0 \ \textrm{for any} \ \Lambda<0.
\label{eqA18c}
\end{align}
\end{subequations}
Here, one of the species $W$ or $U$ may be immobile so that either $\sigma_{w}=0$ or $\sigma_{u}=0$.

Thus, if conditions~(12),~(13) or~(14) are satisfied, the oscillatory Turing instability will be found in the respective ecological networks
as the mobilities $\sigma_{u}, \sigma_{v}$ or $\sigma_{w}$ are increased.

Note that the prey death rates $R$ should be increasing functions of the predator densities.
Hence, in the food chain (Fig.~\ref{fig01}(a)), we have $\partial R^{u} / \partial v > 0$ and $\partial R^{v} / \partial w > 0$ and therefore $F_{v} = \partial F / \partial v <0$ and $G_{w} = \partial G / \partial w <0$.
Moreover, the predator reproduction rates should generally increase with the prey densities.
This implies that $\partial Q^{v} / \partial u > 0$ and $\partial Q^{w} / \partial v > 0$ so that we have $G_{u}=\partial G / \partial u > 0$ and $H_{v}=\partial H / \partial v > 0$.
Thus, the following inequalities are satisfied.
\begin{align}
F_{v} <0, \ F_{w} =0, \ G_{w} <0, \ G_{u}>0, \ H_{w}=0, \ H_{v}>0.
\label{eqS02}
\end{align}

In the food web shown in Fig.~\ref{fig01}(b), both species $V$ and $W$ play a role of the predators for prey $U$ while $V$ is also a prey for $W$.
Therefore, $Q^{u}=Q^{u}(u), R^{u}=R^{u}(u,v,w), Q^{v}=Q^{v}(u,v), R^{v}=R^{v}(v,w), Q^{w}=Q^{w}(u,v,w)$ and $R^{w}=R^{w}(w)$.
Hence, we have $\partial R^{u} / \partial v > 0, \partial R^{u} / \partial w > 0, \partial R^{v} / \partial w > 0, \partial Q^{v} / \partial u > 0, \partial Q^{w} / \partial u > 0$
and $\partial Q^{w} / \partial v > 0$.
This leads to the conditions
\begin{align}
F_{v} <0, \ F_{w} <0, \ G_{w} <0, \ G_{u}>0, \ H_{u}>0, \ H_{v}>0.
\label{eqS03}
\end{align}

In the food web shown in Fig.~\ref{fig01}(c), species $U$ and $V$ are the prey for predator $W$.
Now, we have $Q^{u}=Q^{u}(u), R^{u}=R^{u}(u,w), Q^{v}=Q^{v}(v), R^{v}=R^{v}(v,w), Q^{w}=Q^{w}(u,v,w)$ and $R^{w}=R^{w}(w)$.
Therefore $\partial R^{u} / \partial w > 0, \partial R^{v} / \partial w > 0, \partial Q^{w} / \partial u > 0$ and $\partial Q^{w} / \partial v > 0$.
This implies 
\begin{align}
F_{v} =0, \ F_{w} <0, \ G_{w} <0, \ G_{u}=0, \ H_{u}>0, \ H_{v}>0.
\label{eqS04}
\end{align}
Thus, in all considered food webs, the inequalities
\begin{align}
F_{v}G_{u}\leq 0,\quad G_{w}H_{v}\leq 0 \quad \textrm{and} \quad H_{u}F_{w}\leq 0
\label{eqS16}
\end{align}
hold and at least one of the inequalities holds strictly.
We notice that they imply that the conditions~(\ref{eqA16b}),~(\ref{eqA17b}), and~(\ref{eqA18b}) are all satisfied.
Therefore, we can conclude that, in contrast to general reaction-diffusion models, at least one of the three sufficient conditions will be {\it always satisfied} for ecological networks with three species.
Hence, the oscillatory Turing instability should be more common in ecology, as compared with chemical systems.

\section{Numerical investigations}
\label{sec_num}
\subsection{Setup of numerical simulation}
\label{sec_setup}
Below we numerically investigate the oscillatory Turing instability in ecological networks.
The 4th-order Runge-Kutta scheme with time step $\Delta t = 10^{-3}$ was employed in numerical integration.
The simulations were started from the uniform steady state $(u_{0}, v_{0}, w_{0})$ with random small perturbations with standard deviations $(u_{0}, v_{0}, w_{0})\times 10^{-3}$.
As a typical example of network architecture, we employed scale-free networks generated by the Barab\'{a}si-Albert preferential attachment algorithm~\cite{Barabasi1999}.
The network size was $N=50$ and the mean degree was $\langle k \rangle = 8$.
For convenience, the nodes $i$ were sorted in the decreasing order of their degrees $k_{i}=\sum_{j}A_{ij}$ so that we had
$k_{1}\geq k_{2}\geq \cdots \geq k_{N}$.
To quantify emergent patterns, variation amplitudes for individual nodes $i$ at time $t$ were used;
\begin{align}
A_{i}(t) = \sqrt{\left(u_{i}(t)-\langle u(t) \rangle \right)^{2} + \left(v_{i}(t)-\langle v(t) \rangle \right)^{2} + \left(w_{i}(t)-\langle w(t) \rangle \right)^{2}}
\label{eq03}
\end{align}
where
\begin{align}
& \langle u(t) \rangle = \sum_{i}u_{i}(t)/N,\cr
& \langle v(t) \rangle = \sum_{i}v_{i}(t)/N,\label{eq05}\\
& \langle w(t) \rangle = \sum_{i}w_{i}(t)/N\nonumber
\end{align}
are the mean network quantities.
Time-averaged amplitudes were computed as
\begin{align}
\bar A_{i} = \frac{1}{T}\int_{0}^{T}dt A_{i}(t)
\label{eq04}
\end{align}
after discarding the transients with the averaging time $T=100000$. 

\subsection{Turing instabilities in ecological networks}
\label{sec_modelA}
We first show the results for Model A.
The parameters in Eq.~(\ref{eq02}) are fixed at $a_{u} = 1, b_{u} = 1, c_{u} = 1, a_{v} = 1/4, c_{v} = 1,d_{v} = 1,a_{w} = 1/2,d_{w} = 1$ and $\mu=\nu=1/2$,
yielding a uniform stationary state $(u_{0},v_{0},w_{0}) = (1/2,1/2,1/4)$.
Jacobian matrix at the steady state is
\begin{equation}
J=
\begin{pmatrix}
-1/4 & -1/2 & 0\\
1/4 & 1/8 & -1/2\\
0 & 1/8 & 0
\end{pmatrix}
,
\label{eq_Jacob}
\end{equation}
which satisfies the sufficient conditions (13) irrespective of the mobilities $\sigma_{v}$ and $\sigma_{w}$ of the two species $V$ and $W$.
Then, starting from equal mobilities $\sigma_{u}=\sigma_{v}=\sigma_{w}$,
we gradually increase the mobility $\sigma_{u}$ of the bottom prey $U$ and find the oscillatory Turing instability.

\begin{figure}[tb]
\includegraphics[width=155mm]{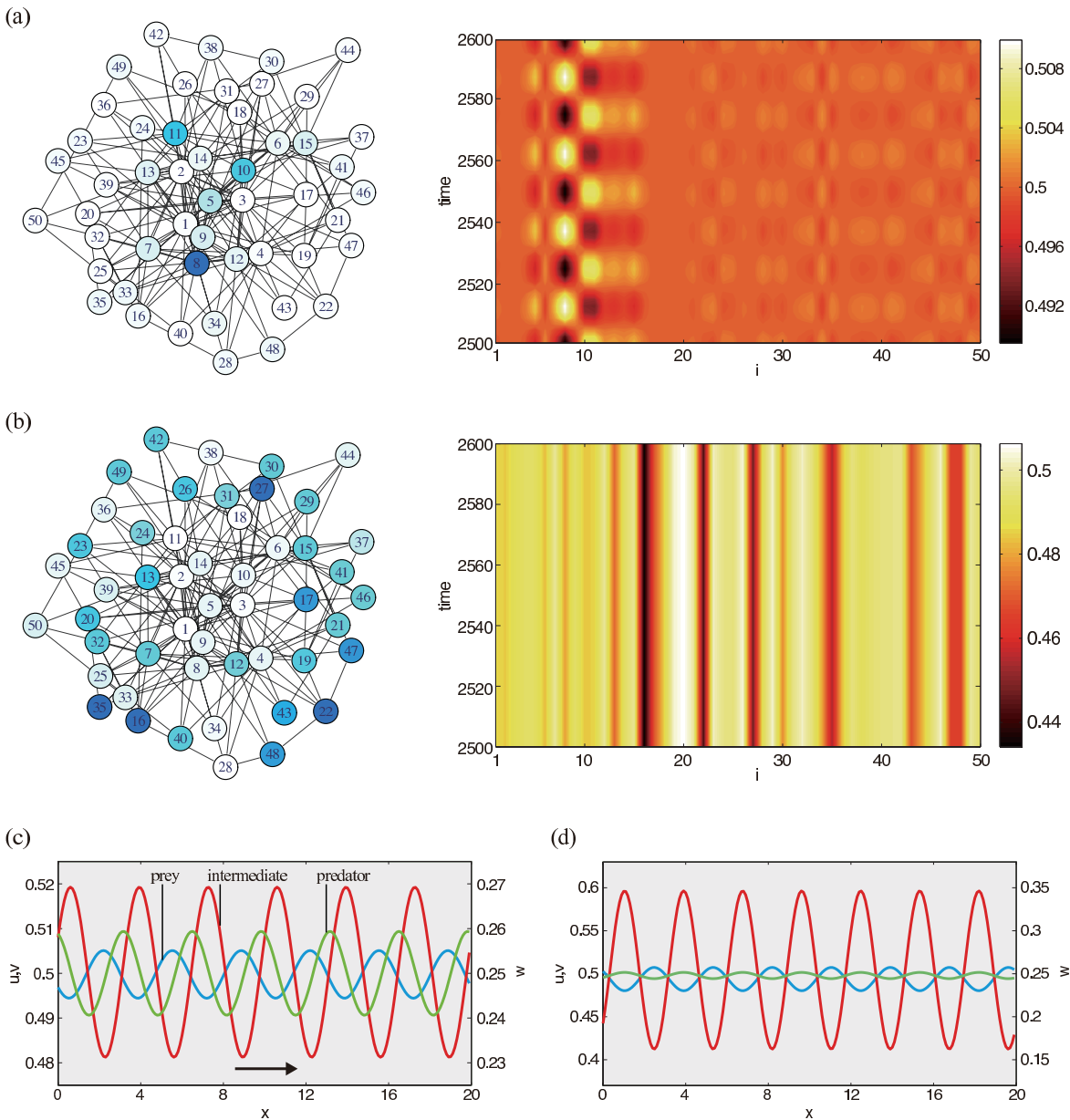}
\caption{
Examples of oscillatory (a,c) and stationary (b,d) Turing patterns in networks (a,b) and in the respective continuous media (c,d).
In panels (a) and (b), mean oscillation amplitudes for all network nodes (see Methods, higher amplitudes are indicated by the deeper blue color) and time evolution diagrams are displayed.
For continuous media, instantaneous concentration profiles are shown (c,d).
The parameters are (a) $\sigma_{u}= 0.535, \sigma_{v}= \sigma_{w}=0.01$, $\epsilon = 0.22$,
(b) $\sigma_{u}= \sigma_{w}=1, \sigma_{v}= 0.022$, $\epsilon = 0.5$,
(c) $\sigma_{u}= 0.535, \sigma_{v}= \sigma_{w}=0.01$, $\epsilon' = 0.8$,
(d) $\sigma_{u}= \sigma_{w}=1, \sigma_{v}= 0.022$, $\epsilon' = 0.65$.
}
\label{fig02}
\end{figure}

The stationary steady state becomes unstable above a certain threshold and non-uniform oscillations develop.
The emerging oscillatory Turing pattern is displayed in Fig.~\ref{fig02}(a).
One can notice that oscillations in the pattern are localized on a subset of nodes.
Network nodes with relatively high degrees (hubs) exhibit oscillations while the other nodes remain staying near the uniform steady state.
We will show later that such localized oscillations are typical for the oscillatory Turing patterns in networks.
Remarkably, the stationary Turing instability can also be found in the same model when the mobility $\sigma_{v}$ of the intermediate species $V$ is decreased (Fig.~\ref{fig02}(b)).

Numerical simulations of the continuous model yield the behavior shown in Figs.~\ref{fig02}(c)(d), where periodic boundary conditions are employed.
The stationary pattern (Fig.~\ref{fig02}(b)) becomes transformed to a periodic stationary Turing pattern (Fig.~\ref{fig02}(d)),
whereas the localized oscillatory pattern (Fig.~\ref{fig02}(a)) gives rise to a traveling wave (Fig.~\ref{fig02}(c)).
While patterns in networks and continuous media may appear different,
they indeed correspond to the same, stationary or oscillatory, Turing bifurcations.

\begin{figure}[t]
\begin{center}
\includegraphics[width=105mm]{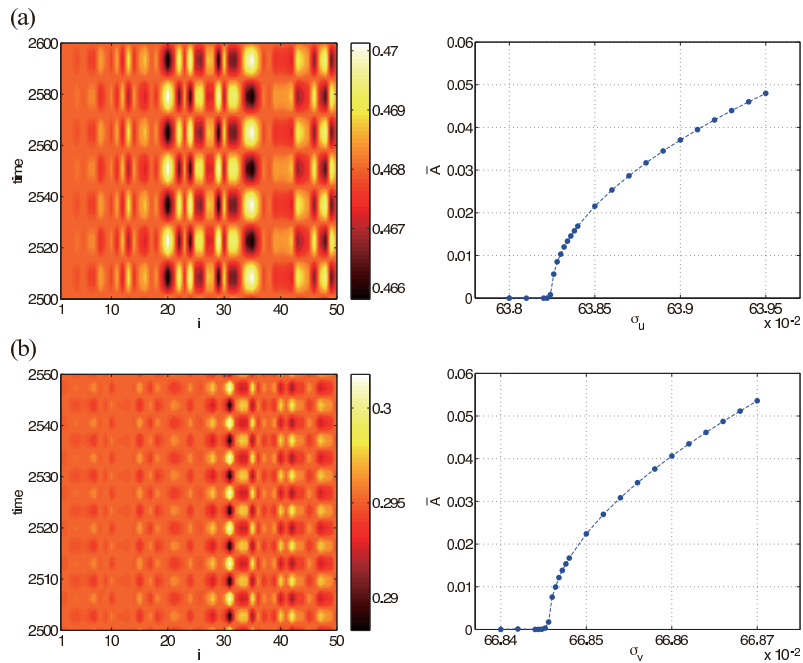}
\caption{
Oscillatory Turing instabilities in Model B and C.
Left panels show final oscillatory Turing patterns.
Right panels show the global amplitude $\bar A$ as a function of $\sigma_{u}$.
Dispersal mobilities are 
(a)
$\sigma_{v} = \sigma_{w}=0.01$ and $\epsilon= 0.25$.
(b)
$\sigma_{v} = \sigma_{w}=0.01$ and $\epsilon = 1.6$.
For the final patterns, (a) $\sigma_{u}= 0.6384$ and (b) $\sigma_{u}= 0.6685$ are used.
}
\label{figA01}
\end{center}
\end{figure}

The oscillatory Turing instability is observed also in other food webs, Models B (Eqs.~(\ref{eqS06})) and C (Eqs.~(\ref{eqS07})).
For Model B, we fixed parameters in Eqs.~(\ref{eqS06}) at
$a_{u}= 1, b_{u}= 1.5, c_{u}= 1, e_{u}= 0.4, a_{v}= 0.25, c_{v}= 1, d_{v}= 1, a_{w}= 0.5, d_{w}=1, e_{w}= 0.4$ and $\mu=\nu=\rho=0.5$,
which gives a uniform steady state $(u_{0},v_{0},w_{0}) \simeq (0.468, 0.221, 0.168)$.
Parameters for Model C are fixed at
$a_{u}= a_{v}=5.55, b_{u}=b_{v}= 1.2, c_{u}=d_{v}=3, a_{w}= 1, c_{w}=1.5, d_{w}=1.5$ and $\mu=\nu=2$ in Eqs.~(\ref{eqS07}),
yielding a uniform steady state $(u_{0},v_{0},w_{0}) \simeq (0.295, 2.330, 3.975)$.
As well as in Model A, the instability takes place in each system when the mobility of one species is increased to exceed a certain threshold.
In Model B, the oscillatory Turing instability occurs when the mobility $\sigma_{u}$ of the bottom prey $U$ is increased.
The emerging pattern is shown in the left panel of Fig.~\ref{figA01}(a).
In Model C, the instability occurs as the mobility $\sigma_v$ of a prey $V$ is increased up (Left panel in Fig.~\ref{figA01}(b)).
Thus, the oscillatory Turing instability is observed in all possible food webs with three species shown in Fig.~\ref{fig01}(a)-(c).

Furthermore, the oscillatory Turing instablity is not sensitive to the choice of nonlinear functions in Eqs.~(\ref{eq01}).
As shown in Supporting Information, different nonlinear functions can be used to describe predator-prey interactions.
We can observe the oscillatory Turing instability under various assumptions~\cite{Murray2003, Holling1959} about the functional form of the reproduction and death rates.
Moreover, the instability can be observed even if one of the three species was immobile (see Supporting Information).

Therefore, we found the oscillatory Turing instability in a wide range of ecological models and conclude that the oscillatory Turing instability is {\it generic} for ecosystems.
This discovery agrees with the statement of our analytical investigation (Sec.~\ref{sec_sufficient}).
Examining emergent patterns, one can notice that developing oscillations in all considered systems are localized on a subset of network nodes with close degrees.
Although the localizing nodes are different depending on a system, localized oscillations were always found.
As we discuss below, localization is a common characteristic of oscillatory Turing patterns in ecological networks.

\subsection{Subcritical vs. supercritical bifurcations}
\label{sec_critical}

Above the instability threshold, nonlinear effects become important.
They can lead to the saturation of growth and the establishment of a final pattern.
Generally, they also determine whether a bifurcation is subcritical or supercritical.
When the bifurcation is subcritical, the pattern with large magnitude becomes immediately established once the instability threshold is exceeded.
Such bifurcations are characterized by hysteresis, so that the pattern persists even below the instability threshold.
In contrast to this, a supercritical bifurcation does not show a hysteresis and the magnitudes of established patterns are small close to the threshold.
In this case, the final patterns do not differ much from the first critical modes near the bifurcation point.

To investigate the bifurcation behavior of the oscillatory and stationary Turing instabilities shown in Fig.~\ref{fig02}, amplitudes of developing patterns can be examined.
The amplitude $\bar A$ of a final pattern on a network can be defined as
\begin{equation}
\bar A = \frac{1}{T} \int_{0}^{T}dt \sqrt{\sum_{i=1}^N \left [ \left(u_{i}(t)-\langle u(t) \rangle \right)^{2} + \left(v_{i}(t)-\langle v(t) \rangle \right)^{2} + \left(w_{i}(t)-\langle w(t) \rangle \right)^{2} \right ]},
\label{eq13}
\end{equation}
where the network averages $\langle u(t) \rangle,\langle v(t) \rangle$ and $\langle w(t) \rangle$ were defined in Eqs.~(\ref{eq05}).
Note that this definition holds both for the oscillatory and stationary patterns.

\begin{figure}[t]
\begin{center}
\includegraphics[width=105mm]{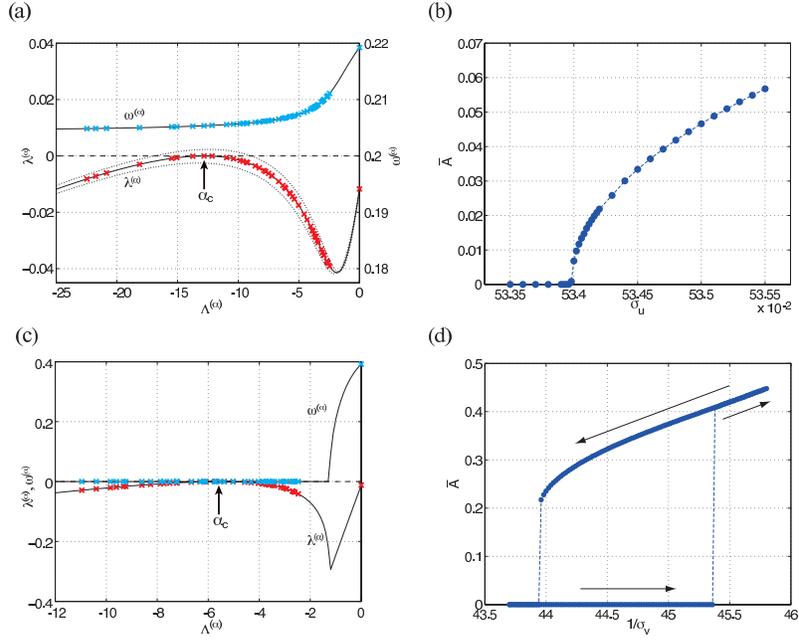}
\caption{
Linear stability analysis on Model A.
Growth rates $\lambda^{(\alpha)}$ and frequencies $\omega^{(\alpha)}$ for different modes $\alpha$ near the instability threshold (a,c) and the global amplitudes $\bar A$ (b,d)
for oscillatory (a,b) and stationary (c,d) Turing instabilities.
The arrows in (d) show the directions of the parameter change.
Dispersal mobilities are
(a) $\epsilon = 0.45, \sigma_{u}=0.5342, \sigma_{v}=\sigma_{w}=0.01$,
(b) $\epsilon = 0.45, \sigma_{v}=\sigma_{w}=0.01$,
(c) $\epsilon = 0.5, \sigma_{v}=0.022, \sigma_{u}=\sigma_{w}=1$,
(d) $\epsilon = 0.5, \sigma_{u}=\sigma_{w}=1$.
}
\label{fig03}
\end{center}
\end{figure}

Figure~\ref{fig03} displays the results of the linear stability analysis and the global amplitude $\bar A$ for the oscillatory (a)(b) and stationary (c)(d) Turing instabilities in Model A.
Increasing the mobility $\sigma_{u}$ of the prey $U$, we calculated the linear growth rate $\gamma = \lambda + i\omega$
to find the instability at a threshold $\sigma_{u, \textrm{crit.}}$.
The critical eigenmode has a complex linear growth rate (Fig.~\ref{fig03}(a)) and therefore the instability is oscillatory.
After passing the instability, we gradually increase the mobility $\sigma_{u}$ further away from the threshold
and calculate the global amplitude $\bar A$ (Fig.~\ref{fig03}(b)).
As can be clearly seen in the figure, the oscillatory Turing instability corresponds to a supercritical bifurcation.
We have checked that near the threshold $\bar A \propto (\sigma_{u}-\sigma_{u, \textrm{crit.}})^{1/2}$ holds.
Thus, small amplitude patterns can be established near the instability threshold.
In contrast, the stationary Turing instability, which corresponds to a growth rate of a real number (Fig.~\ref{fig03}(c)), exhibits a subcritical bifurcation.
As shown in Fig.~\ref{fig03}(d), once the instability takes place, the amplitude $\bar A$ jumps up to a large value.

\subsection{Estimations of oscillatory Turing patterns using critical Laplacian eigenvectors.}
\label{sec_esti}

\begin{figure}[t]
\includegraphics[width=150mm]{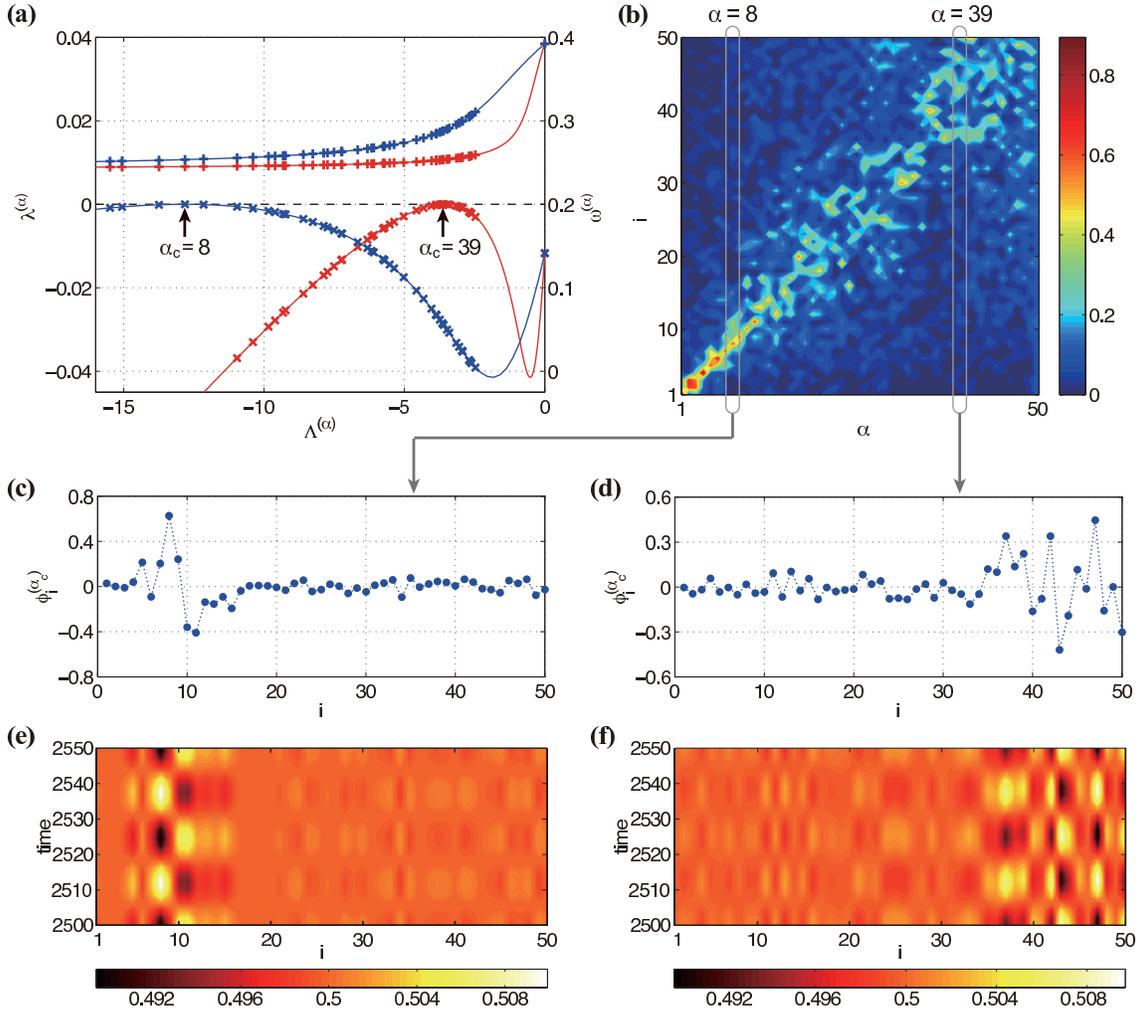}
\caption{
Oscillatory Turing instability in Model A.
(a) Dependences of $\lambda^{(\alpha)}$ and $\omega^{(\alpha)}$ on the Laplacial eigenvalue $\Lambda^{(\alpha)}$ for different overall mobilities, $\epsilon = 0.22$ (red curves) and $\epsilon = 0.79$ (blue curves);
relative mobilities are $\sigma_{u}= 0.535, \sigma_{v}= \sigma_{w}=0.01$ for both cases.
(b) Laplacian spectrum of the network is graphically displayed.
Each column corresponds to one eigenvector $\phi^{(\alpha)}$.
Nodes are sorted according to their degrees $k$ and the magnitude $\left | {\phi}_{i}^{(\alpha)} \right |$ for each node $i$ is indicated by using a color code.
(c-f) Critical Laplacian eigenvectors with (c) $\alpha=8$ and (d) $\alpha=39$ and the respective final oscillatory Turing patterns (e,f).
}
\label{fig04}
\end{figure}

Due to its supercritical bifurcation,
oscillatory Turing patterns near the instability threshold are well described only by the first critical eigenmode $\alpha_{\textrm c}$, so that we have
\begin{align}
& \delta u_{i}(t) = U^{(\alpha_{\textrm{c}})}\exp [\omega^{(\alpha_{\textrm{c}})} t] \phi^{(\alpha_{\textrm{c}})}_{i}\cr
& \delta v_{i}(t) = V^{(\alpha_{\textrm{c}})}\exp [\omega^{(\alpha_{\textrm{c}})} t] \phi^{(\alpha_{\textrm{c}})}_{i}\label{eq14}\\
& \delta w_{i}(t) = W^{(\alpha_{\textrm{c}})}\exp [\omega^{(\alpha_{\textrm{c}})} t] \phi^{(\alpha_{\textrm{c}})}_{i}.\nonumber
\end{align}
Therefore, in this case, final patterns can be predicted by means of the critical Laplacian eigenvectors $\vec{\phi}^{(\alpha_{\textrm{c}})}$.

In Fig.~\ref{fig04}, we demonstrate the prediction for Model A.
Figure~\ref{fig04}(a) shows the results of the linear stability analysis at two different values of overall mobility $\epsilon$ in Model A.
The respective critical Laplacian eigenvectors are displayed in Figs.~\ref{fig04}(c)(d) and the actual oscillatory Turing patterns in Figs.~\ref{fig04}(e)(f).
As seen in Fig.~\ref{fig04}, critical Laplacian eigenvectors and developing oscillatory patterns are localized on subsets of network nodes with close degrees.
The localization for the considered scale-free network of size $N=50$ is demonstrated in Fig.~\ref{fig04}(b),
where magnitudes $\left | \phi_{i}^{(\alpha)}\right |$ of the components of Laplacian eigenvectors $\vec{\phi}^{(\alpha)}$ are displayed as a function of $\alpha$.
As we explained in Sec.~\ref{sec_linear_stab}, the mode indices $\{\alpha\}$ are sorted in the increasing order of the Laplacian eigenvalues $\{ \Lambda^{(\alpha)} \}$ so that $\Lambda^{(1)}\leq \Lambda^{(2)}\leq \cdots \leq \Lambda^{(N)} = 0$ holds.
This localization is consistent with the previous discovery for large random networks~\cite{Nakao2010, McGraw2008}.
The critical eigenmode $\alpha_{\textrm{c}}$ depend on the overall mobility $\epsilon$ shown in Fig.~\ref{fig04}(a).
As discussed previously~\cite{Nakao2010}, the Laplacian eigenvalue $\Lambda^{(\alpha)}$ appears in the characteristic function~(\ref{eq10}) multiplied by $\epsilon$.
Then, if we change the overall diffusion mobility $\epsilon$, the critical eigenmode $\alpha_{\textrm c}$ shifts so that the product $\epsilon \Lambda^{(\alpha_{\textrm c})}$ is kept constant.
Thus, the characteristic localization of the critical Laplacian eigenvector changes depending on the overall mobility $\epsilon$ shown in Fig.~\ref{fig04}(c)(d).
Correspondingly, in emerging oscillatory Turing patterns, localizing nodes shift from hubs (Fig.~\ref{fig04}(e)) to peripheral nodes (Fig.~\ref{fig04}(f)).
Thus, examining Fig.~\ref{fig04}, one can predict the emerging patterns by means of respective critical Laplacian eigenvectors.
%

The oscillatory Turing instabilities found in other models (Models B and C, and other models shown in Supporting Information) are also identified with supercritical bifurcations shown in Figs.~\ref{figA01},~\ref{figA02} and~\ref{figA03}.
Thus, the instability generically leads to pattern formations of small amplitudes, which are described by the Laplacian eigenvectors of critical modes.
This indicates that the localization property of the oscillatory patterns originates in the localizing Laplacian eigenvectors. 

In contrast to the oscillatory instability, the stationary Turing instability corresponds to a subcritical bifurcation.
This gives a striking difference between patterns arising from both the instabilities.
It has been previously found in two-component reaction-diffusion networks that
the stationary Turing instability always corresponds to a subcritical bifurcation~\cite{Nakao2010}.
This holds also in three-component systems, that is, ecological networks with three species.
The bifurcation is subcritical and the stable nonlinear state is far from the uniform steady state.
Although differentiation starts from the nodes with the characteristic degree of the critical eigenvector,
it takes place in surrounding nodes sequentially and spreads over the entire network.
The amplitude of the developed final pattern becomes large.
Thus, as we have seen in Fig.~\ref{fig02},
although both instabilities are induced by the diffusion effect,
the uniform steady state is destabilized in distinctly different fashions.

\subsection{Secondary instability - Extinction of ecological species}
\label{sec_extinction}
\begin{figure}[t]
\includegraphics[width=120mm]{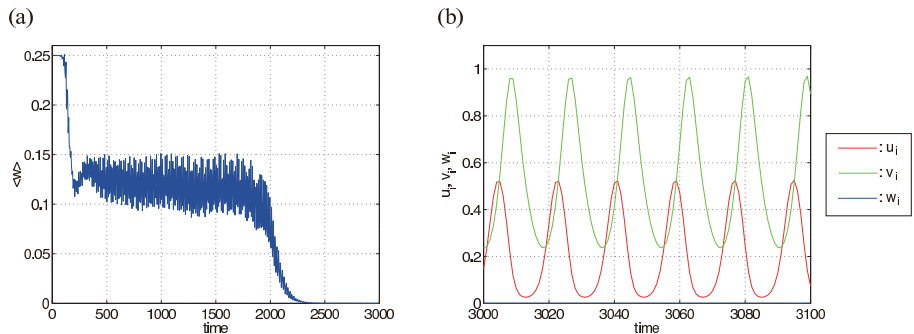}
\caption{
Secondary instability in Model A.
(a) Time-dependence of the network average $\langle w(t) \rangle$ far from the instability threshold.
(b) Local densities of three species after a transient.
The same dynamics are observed in all network nodes
(Uniform oscillations).
Dispersal mobilities are $\epsilon = 0.1, \sigma_{u}=6.75, \sigma_{v}=\sigma_{w}=0.01$.
}
\label{fig06}
\end{figure}

We have studied the dynamical behavior induced by the oscillatory Turing instability far from the threshold
and found a potential secondary instability, leading to the extinction of ecological species.
Note that the dynamical behavior far from the instability threshold largely depends on the considered model.
We have identified such extinction in Models A and B.
As an illustration, we here show the results for Model A.
Figure~\ref{fig06} displays numerical results far from the first instability threshold.
Now, all eigenmodes are unstable except for the zero eigenmode $\alpha = 0$ (Hopf mode).
In panel (a), the network average $\langle w(t) \rangle$ for the top predator $W$ is plotted as a function of time.
After an oscillatory transient, the average $\langle w(t) \rangle$ tends to zero and therefore the top predator vanishes.
Two remaining species, the prey $U$ and the intermediate predator $V$, exhibit uniform oscillations after the extinction of the top predator $W$ (Fig.~\ref{fig06}(b)).
Thus, the secondary instability may be inherent in the oscillatory Turing instability, which destabilizes the considered ecosystem and leads to the extinction of ecological species.

\section{Discussion and conclusions}
\label{sec_sum}
The mathematical description for the classical (stationary) Turing instability in networks has been proposed already in 1971~\cite{Othmer1971}.
However, it has been first applied only to regular lattices~\cite{Othmer1971, Othmer1974} and small networks~\cite{Horsthamke2004, Moore2005}.
Recently, such instability in large random networks has been investigated and characteristic properties of stationary Turing patterns in large networks have been discussed~\cite{Nakao2010, Hata2012, Wolfrum2012}.
In contrast to this, the mathematical theory of the oscillatory Turing bifurcation (the analogue of the wave bifurcation) in networks has been missing and our work is the first report where such theory is constructed.

Similar to continuous reaction-diffusion systems, the oscillatory instability needs three reacting species~\cite{Turing1952}
and it occurs when diffusion mobilities of the species are largely different.
However, wave patterns, which are typical for such instability in continuous media, do not emerge in networks.
Therefore, we prefer not to use the term ``wave instability'' in the present study.
As we find, heterogeneous oscillations spontaneously develop and they are localized on a subset of network nodes with similar degrees.
The localization of developing oscillations could be explained by taking into account the known statistical properties of Laplacian eigenvectors in large random networks~\cite{Nakao2010, McGraw2008}.
Such eigenvectors correspond to the critical modes of the oscillatory Turing instability
and hence they determine the properties of developing patterns.

Previously, it has been shown that the classical (stationary) Turing bifurcation in networks is always subcritical; it is characterized by strong hysteresis and the properties of final stationary patterns in networks are largely different from those of the first critical modes.
In contrast to this, we have found that the oscillatory Turing bifurcation in networks is typically supercritical.
Thus, the hysteresis is absent and the amplitude of the Turing patterns gradually grows as the control parameter is increased.
Near the bifurcation point, oscillatory Turing patterns with small amplitudes are observed and they agree well with the first critical modes.
This means that by considering Laplacian eigenvectors the properties of final small-amplitude patterns can be analyzed.

While the proposed mathematical theory is general and applicable to systems with various origins,
our detailed numerical simulations have been performed for the models which correspond to ecological networks also known as metapopulations~\cite{Hanski1998}.
We have considered metapopulations with the graph structure of scale-free networks.
The habitats, representing network nodes, were occupied by local three-species ecosystems forming food webs.
All possible food webs with three predator or prey species were considered.
We have only excluded the food web with two predators and one prey shown in Fig.~\ref{fig01} (d) 
because steady coexistence of all three species is not possible in this case.
Moreover, simulations under various assumptions for nonlinear predator-prey interactions have been carried out.

The oscillatory Turing instability could be observed for all considered metapopulation models.
It has been found as the network mobility (dispersal rate) of one of the species was gradually increased.
Note that it could be the mobility of a prey or a predator, depending on the food web and the mathematical model applied.
The instability has been observed also in the three-species metapopulations where one of the species was immobile.
The results of our numerical simulations are supported by the general sufficient conditions for the oscillatory bifurcation in three-component ecological systems which we have derived.
While our analysis has been performed only for systems with three species, it can be straightforwardly extended to the systems with a larger number of components and hence for the metapopulations with more complex food webs.

We conclude that the oscillatory Turing instability may be {\it generic} for ecosystems.
It should be generally expected whenever metapopulations with at least three species and sufficiently large differences in the mobilities of the species are investigated.
This result may be of principal importance.
Previously, the discussion has been focused on the role of dispersal connections in enhancing the stability of uniform steady states~\cite{Hanski1991,Gonzalez1998}.
We find however that dispersal connections would often destabilize the uniform steady state
and lead to the development of oscillations on a subset of network nodes.
We would like to stress that such oscillations are a consequence of the differential dispersal mobilities of species and they were always absent for isolated populations in individual habitats. 

While the developing oscillations are weak near the instability threshold,
their amplitude increases with the distance from the critical point
and large-amplitude oscillations are also possible.
We have shown that the secondary instabilities of such oscillations may take place and, in the considered example Fig.~\ref{fig06}, they have resulted in the extinction of one of the species and, therefore, in the degradation of an ecosystem.
So far, only the global extinction through the development of uniform oscillations via a Hopf bifurcation has been discussed.
Our work provides a different scenario for the extinction of species through the development of dispersal-induced hegerogeneous oscillations in ecological networks.

It would be interesting to perform experimental or field studies of the oscillatory Turing instability
and the resulting Turing patterns in real ecological systems.
For this purpose, it may be beneficial to work first with the artificially constructed metapopulations as it has been done before in the experimental studies on the role of dispersal connections~\cite{Gonzalez1998, Keymer2006, Balagadde2008}.
To prove the presence of oscillatory Turing patterns in natural ecosystems, the development of such patterns and their responses to local perturbations should be analyzed, similar to what has been done to demostrate the existence of classical stationary Turing patterns in biological organisms~\cite{Kondo1995}.

\section{Appendices}
\label{sec_app}

\subsection{Numerical simulations of other ecological models}
\begin{figure}[t]
\begin{center}
\includegraphics[width=105mm]{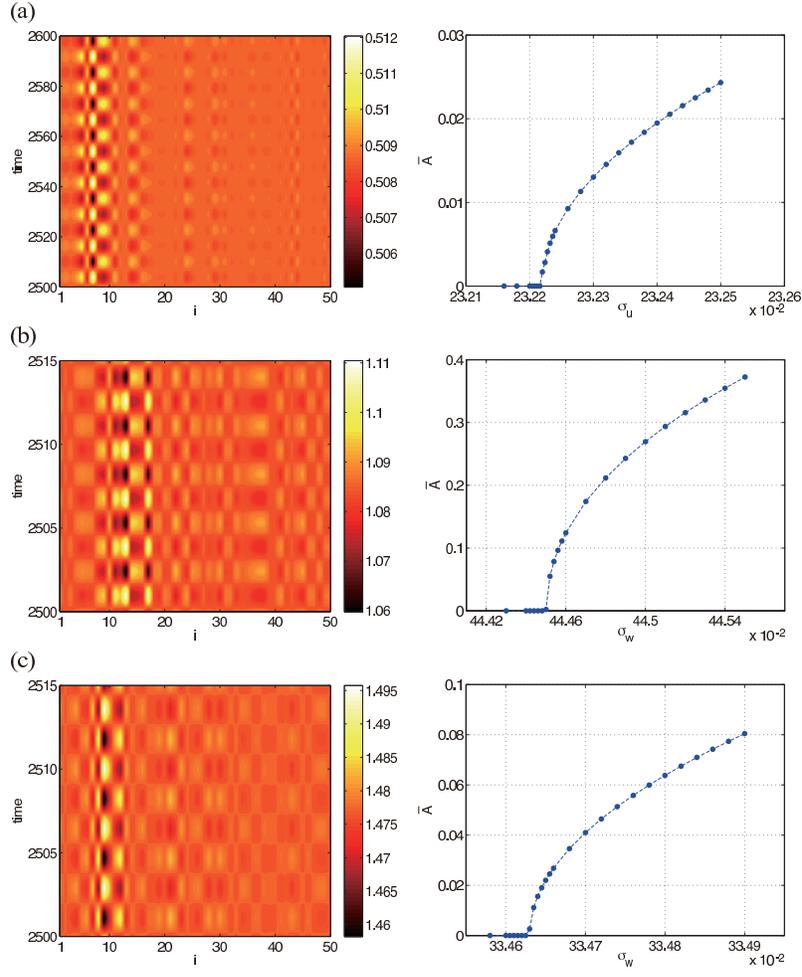}
\caption{
Oscillatory Turing instabilities in 
(a) Model D, (b) Model E and (c) Model F.
Left panels show final oscillatory Turing patterns.
Right panels show the amplitude $\bar A$ as a function of $\sigma_{u}$ or $\sigma_{w}$.
Dispersal mobilities are fixed at
(a) $\sigma_{v} = \sigma_{w}=0.01$ and $\epsilon= 0.34$.
(b) $\sigma_{u} = \sigma_{v}=0.01$ and $\epsilon= 1.6$.
(c) $\sigma_{u} = \sigma_{v}=0.01$ and $\epsilon= 1.6$.
For the final patterns, (a) $\sigma_{u}= 0.2323$, (b) $\sigma_{w}= 0.4446$ and (c) $\sigma_{w}= 0.3347$ are used.
}
\label{figA02}
\end{center}
\end{figure}

Here we show the results for different ecological models.
Different dependence of reproduction and death rates on the densities of the species in predator-prey models are possible~\cite{Murray2003}.
In the main text, Holling type II dependences have been used as a typical example.
Below we give models with other dependences including Holling type III~\cite{Holling1959, Murray2003} and the functions introduced by Murray~\cite{Murray2003}.
As the food web architecture, the food chain shown in Fig.~\ref{fig01}(a) is employed.\\
{\bf Model D.} The food chain (Fig.~\ref{fig01}(a)) with the Holling type III dependence for both prey and predator:
\begin{align}
&Q^{u}(u) = a_{u} -b_{u}u,& &R^{u}(u,v) = c_{u}\frac{uv}{u^{2}+\mu},\cr
&Q^{v}(u) = c_{v}\frac{u^{2}}{u^{2}+\mu},& &R^{v}(v,w) = a_{v} + d_{v}\frac{vw}{v^{2}+\nu},\label{eqS08}\\
&Q^{w}(v) = d_{w}\frac{v^{2}}{v^{2}+\nu},& &R^{w} = a_{w}.\nonumber
\end{align}
Results are shown in Fig.~\ref{figA02}(a).
Parameters in Eqs.~(\ref{eqS08}) are fixed at
$a_{u}= 2.5, b_{u}= 3, c_{u}= 1.2, a_{v}= 0.2, c_{v}= 1.2, d_{v}= 1, a_{w}= 0.75, d_{w}=1$ and $\mu=\nu=0.125$.
A uniform steady state is found at $(u_{0}, v_{0}, w_{0}) \simeq (0.509, 0.612, 0.497)$.
The oscillatory Turing instability is observed as $\sigma_u$ is increased.\\
{\bf Model E.} The food chain (Fig.~\ref{fig01}(a)) with the Holling type II dependence for prey and a linear function for predator:
\begin{align}
&Q^{u}(u) = a_{u} -b_{u}u,& &R^{u}(u,v) = c_{u}\frac{v}{u+\mu},\cr
&Q^{v}(u,v) = C_{v} + a_{v} \left ( 1 - c_{v}\frac{v}{u} \right ),& &R^{v}(v,w) = C_{v} + d_{v}\frac{w}{v+\nu},\label{eqS09}\\
&Q^{w}(v,w) = C_{w} + a_{w} \left ( 1 - d_{w}\frac{w}{v} \right ),& &R^{w} = C_{w}.\nonumber
\end{align}
Results are shown in Fig.~\ref{figA02}(b).
Parameters in Eqs.~(\ref{eqS09}) are fixed at
$a_{u}= 3, b_{u}= 1, c_{u}= 1, a_{v}= 6, c_{v}= 1/6, d_{v}= 1, a_{w}= 4, d_{w}=0.25$ and $\mu=\nu=0.25$.
A uniform steady state is found at $(u_{0}, v_{0}, w_{0}) \simeq (1.084, 2.557, 10.23)$.
The oscillatory Turing instability takes place in this system as $\sigma_w$ is increased.\\
{\bf Model F.} The food chain (Fig.~\ref{fig01}(a)) with the Holling type III dependence for prey and a linear function for predator:
\begin{align}
&Q^{u}(u) = a_{u} - b_{u}u,& &R^{u}(u,v) = c_{u}\frac{uv}{u^{2}+\mu},\cr
&Q^{v}(u,v) = C_{v} + a_{v} \left ( 1 - c_{v}\frac{v}{u} \right ),& &R^{v}(v,w) = C_{v} + d_{v}\frac{vw}{v^{2}+\nu},\label{eqS10}\\
&Q^{w}(v,w) = C_{w} + a_{w} \left ( 1 - d_{w}\frac{w}{v} \right ),& &R^{w} = C_{w}.\nonumber
\end{align}
Results are shown in Fig.~\ref{figA02}(c).
Parameters in Eqs.~(\ref{eqS10}) are fixed at
$a_{u}= 3, b_{u}= 1, c_{u}= 1.5, a_{v}= 8, c_{v}= 0.25, d_{v}= 2.5, a_{w}= 5, d_{w}=0.4$ and $\mu=\nu=0.25$.
A uniform steady state is $(u_{0}, v_{0}, w_{0}) \simeq (1.477, 1.672, 4.179)$.
The oscillatory Turing instability is found as $\sigma_w$ is increased.

\begin{figure}[tb]
\begin{center}
\includegraphics[width=105mm]{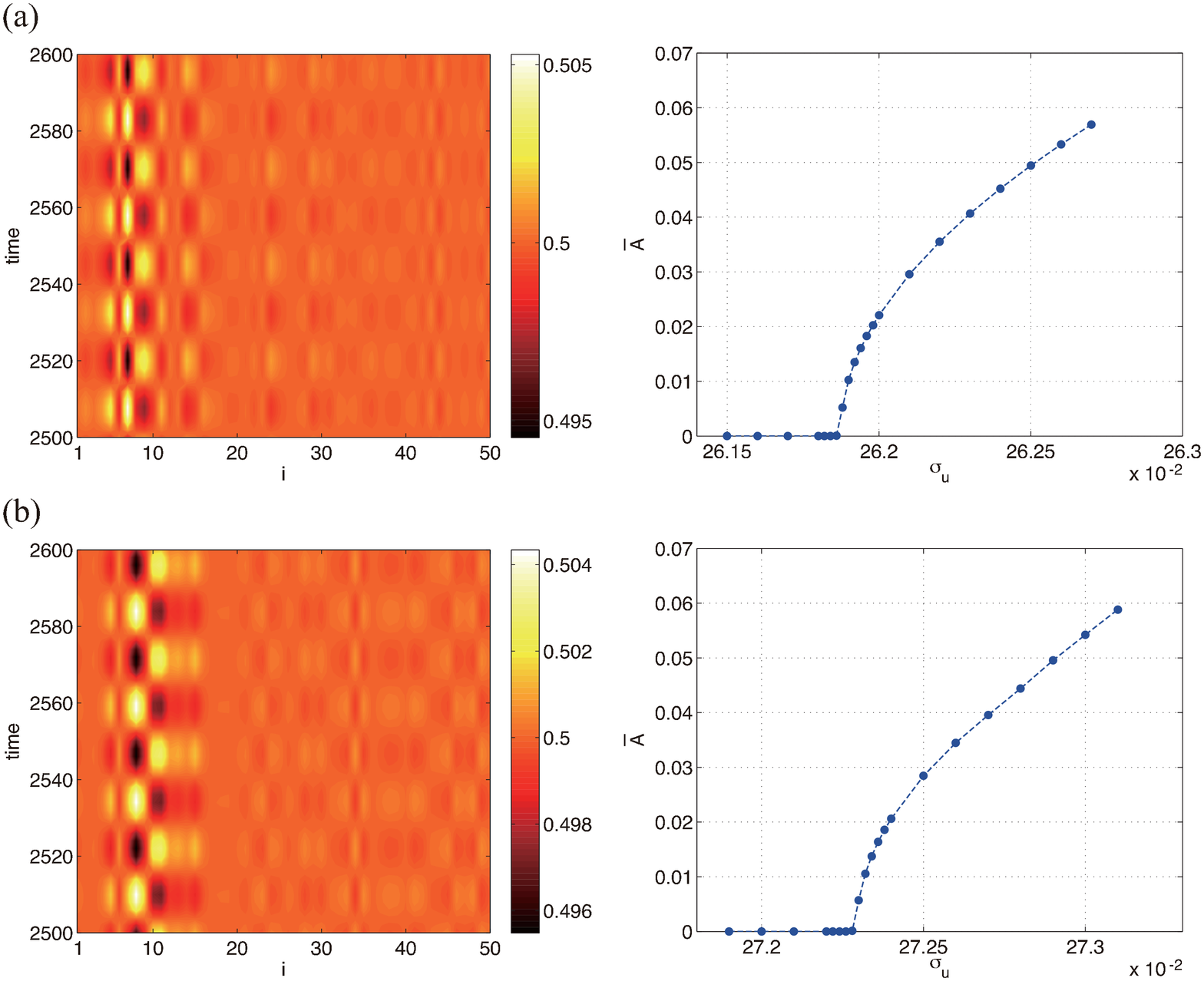}
\caption{
Oscillatory Turing instabilities in food chains (Model A) model with only two mobile species.
Left panels show final oscillatory Turing patterns.
Right panels show the amplitude $\bar A$ as a function of $\sigma_{u}$.
(a) Intermediate predator $V$ is immobile ($\sigma_{v}=0, \sigma_{w}=0.01$).
(b) Top predator $W$ is immobile ($\sigma_{v}=0.01, \sigma_{w}=0$).
Oscillatory Turing patterns in left panels are observed at
(a) $\sigma_{u}= 0.262$
(b) $\sigma_{u}= 0.2724$
with the overall dispersal mobility $\epsilon = 0.4$.
}
\label{figA03}
\end{center}
\end{figure}

Note that the constants $C_{v,w}$ enter both into the reproduction and death rates of species $V$ and $W$ in Eqs.~(\ref{eqS09}) and~(\ref{eqS10}).
Therefore, they become cancelled in the expressions for $G$ and $H$.
Their numerical values are not relevant and are not provided.\\
{\bf Model A with one immobile species.}
As implied by the sufficient conditions (12)-(14), the oscillatory Turing instability is possible even if one of the species is immobile.
Figure~\ref{figA03} shows results for the case that the intermediate species $V$ is immobile, i.e. $\sigma_{v}=0$ (Fig.~\ref{figA03}(a))
and that the top predator $W$ is immobile, i.e. $\sigma_{w}=0$ (Fig.~\ref{figA03}(b)).
Parameters in Eqs.~(\ref{eq02}) are the same as those used in the main text.
The oscillatory Turing instability is observed as $\sigma_u$ is increased in both cases.

Thus, the oscillatory Turing instability could be observed
for different food web architectures (Fig.~\ref{fig03} and~\ref{figA01})
and for different nonlinearities in the reproduction and death rates (Fig.~\ref{figA02}).
Moreover, the oscillatory Turing patterns were observed even when one of the three species was immobile (Fig.~\ref{figA03}).
In all considered systems, oscillatory Turing bifurcations were supercritical.

\subsection{Numerical simulations of chemical reaction-diffusion networks}

In chemical systems, autocatalytic reactions and diffusion may induce the Turing instability.
The chemical Brusselator~\cite{Glandsdorff1971} and the Oregonator~\cite{Field1974} are typical mathematical models to describe autocatalytic reactions
and their numerical investigations are extensively used for exploring pattern formations in chemical systems.
The Brusselator was originaly proposed to examine chemical oscillatory dynamics.
It is a hypothetical model, which gives a rich variety of chemical dissipative structures.
In contrast to the Brusselator, the Oregonator has been proposed to describe the real Belousov-Zhabotinsky reaction.

Both chemical models have been previously extended to explore the oscillatory Turing instability (wave instability) in continuous media~\cite{Zhabotinsky1995, Vanag2001, Yang2002}.
It was suggested that a chemical reaction
in which the activator is reversibly transformed into an unreactive chemical species can explain wave patterns observed in the BZ aerosol OT system~\cite{Vanag2001}.
Therefore, an additional third component taking such reversible transformation was added to the Brusselator.
For the Oregonator, which originally contains three components, an additional reversal reaction was added.
Thus, extended Brusselator and Oregonator were constructed~\cite{Yang2002}.

Here, we consider the behaviors of such extended models in network-organized systems.
The extended network Brusselator is described by equations
\begin{align}
\begin{cases}
\displaystyle \frac{du_{i}}{dt} =  a - (1+b) u_{i} + {u_{i}}^{2}v_{i} -cu_{i}+ dw_{i} + \epsilon \sigma_{u} \displaystyle \sum_{j=1}^N L_{ij} u_j,\\
\displaystyle \frac{dv_{i}}{dt} =  bu_{i} - {u_{i}}^{2}v_{i} + \epsilon \sigma_{v} \displaystyle \sum_{j=1}^N L_{ij} v_j,\\
\displaystyle \frac{dw_{i}}{dt} =  cu_{i} - dw_{i} + \epsilon \sigma_{w} \displaystyle \sum_{j=1}^N L_{ij} w_j.
\label{eqS18}
\end{cases}
\end{align}
We fix parameters as $a=1,b=2.9,c=1$ and $d=1$, yielding a steady state $(u_{0},v_{0},w_{0})=(1,2.9,1)$.
The extended network Oregonator model is given by
\begin{align}
\begin{cases}
\displaystyle \frac{du_{i}}{dt} = \frac{1}{\epsilon_{0}}\left [ u_{i} - {u_{i}}^{2} - p v_{i} \frac{u_{i}-q}{u_{i}+q} -cu_{i}+dw_{i} \right ] + \epsilon \sigma_{u} \displaystyle \sum_{j=1}^N L_{ij} u_j,\\
\displaystyle \frac{dv_{i}}{dt} = u_{i}-v_{i} + \epsilon \sigma_{v} \displaystyle \sum_{j=1}^N L_{ij} v_j,\\
\displaystyle \frac{dw_{i}}{dt} = \frac{1}{\epsilon_{1}} [ cu_{i}-dw_{i} ] + \epsilon \sigma_{w} \displaystyle \sum_{j=1}^N L_{ij} w_j,
\label{eqS19}
\end{cases}
\end{align}
with parameters $p=0.95, q=0.01, c=0.2, d=1, \epsilon_{0}=0.35$ and $\epsilon_{1}=2$,
yielding a steady state $(u_{0},v_{0},w_{0})\simeq(0.161,0.161,0.032)$.

\begin{figure}[tb]
\begin{center}
\includegraphics[width=105mm]{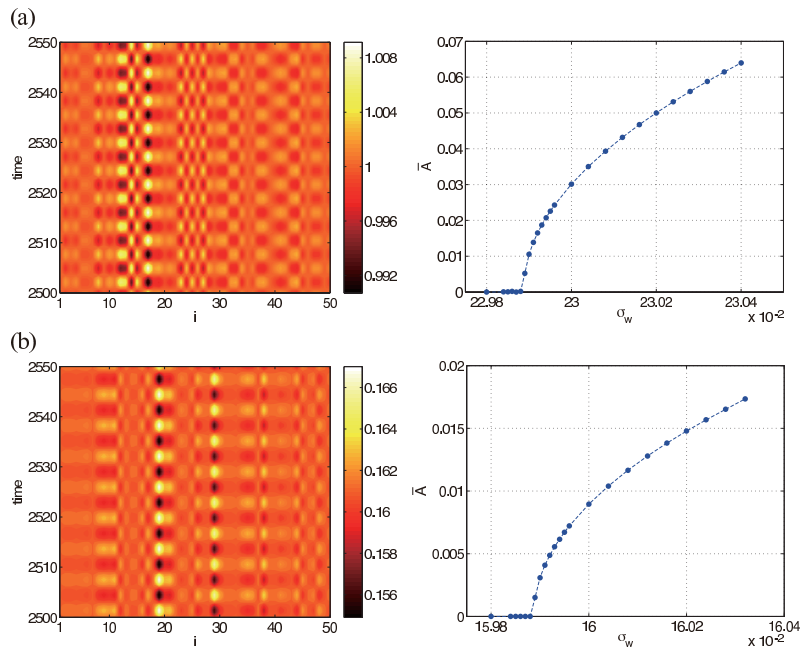}
\caption{
Oscillatory Turing instability in (a) the extended Brusselator model and (b) the extended Oregonator model.
Left panels show final oscillatory Turing patterns.
Right panels show the amplitude $\bar A$ as a function of $\sigma_{w}$.
Diffusional mobilities are fixed at
(a) $\epsilon = 0.35, \sigma_{u}=0.01$ and $\sigma_{v}=0.01$,
(b) $\epsilon = 0.5, \sigma_{u}=0.01$ and $\sigma_{v}=0.01$.
For the final patterns, (a) $\sigma_{w}=0.23$ and (b) $\sigma_{w}=0.16$ are used.
}
\label{figA04}
\end{center}
\end{figure}

Numerical results are shown in Fig.~\ref{figA04}.
In both systems, the oscillatory Turing instability takes place when increasing the diffusion mobility $\sigma_{w}$ of reactant $w$.
The bifurcation is supercritical and leads to localized oscillations in chemical networks.

\section*{Acknowledgements}
Financial support through the DFG  SFB 910 program ``Control of Self-Organizing Nonlinear Systems'' in Germany,
through the Fellowship for Research Abroad, KAKENHI and the FIRST Aihara Project (JSPS),
and the CREST Kokubu Project (JST) in Japan is gratefully acknowledged.

%
%




\end{document}